# Simplifying recombinant protein production: Combining Golden Gate cloning with a standardized protein purification scheme

Sonja Zweng[1], Gabriel Mendoza-Rojas[1], Florian Altegoer[1*]

[1]Institute of Microbiology, Heinrich-Heine University, Düsseldorf, Germany

*Correspondence to: altegoer@hhu.de

## Abstract

Recombinant protein production is pivotal in molecular biology, enabling profound insights into cellular processes through biophysical, biochemical, and structural analyses of the purified samples. The demand for substantial biomolecule quantities often presents challenges, particularly for eukaryotic proteins. *Escherichia coli* expression systems have evolved to address these issues, offering advanced features such as solubility tags, posttranslational modification capabilities, and modular plasmid libraries. Nevertheless, existing tools are often complex, which limits their accessibility and necessitates streamlined systems for rapid screening under standardized conditions. Based on the Golden Gate cloning method, we have developed a simple 'one-pot' approach for the generation of expression constructs using strategically chosen tags like hexahistidine, SUMO, MBP, GST, and GB1 to enhance solubility and expression. Tags are removable via TEV protease cleavage, and the system allows visual cloning verification through mScarlet fluorescence. We provide a comprehensive protocol encompassing oligonucleotide design, cloning, expression, Ni-NTA affinity chromatography, and size-exclusion chromatography. This method therefore streamlines prokaryotic and eukaryotic protein production, rendering it accessible to standard molecular biology laboratories with basic protein biochemical equipment.

**Key words:** Golden Gate, modular cloning, protein expression, solubility tags

## 1. Introduction

Recombinant protein production is one of the cornerstones of modern molecular biology. The characterization of proteins using biophysical, biochemical and structural techniques provides a deep understanding of their functions and enabled to unravel the intricate mechanisms of many cellular processes. Most of these techniques require large quantities of biomolecules, and their production using heterologous systems is often still a bottleneck, especially considering eukaryotic proteins. *E. coli* expression systems have been continuously advanced by e.g. introducing different solubility tags, enhancing the production of post-translationally modified proteins [1], or providing highly modular plasmid libraries for high-throughput screening [2]. Most of these tools are powerful for exhaustive expression screening of difficult-to-produce candidates and therefore likely used in laboratories with extensive biochemical expertise. However, many laboratories lacking this expertise rather require systems that provide easy access and rapid screening using standardized expression conditions.

Various molecular cloning methods are available to clone and assemble DNA fragments of interest. Golden Gate offers a fast and reliable alternative to conventional restriction-insertion cloning relying on type-IIS endonucleases that hydrolyze DNA outside their recognition sequence. This way, multiple fragments can be assembled hierarchically using pre-defined overhangs [3]. Furthermore, the Golden Gate method is highly efficient and thus perfectly suited to assemble constructs in a single 'one-pot' reaction.

Our approach therefore combines a simple collection of Golden Gate-based expression constructs with a streamlined strategy of protein expression conditions. We have included sequences for hexahistidine tags fused N- or C-terminally to the protein of interest (POI), classical solubility enhancing tags as the small ubiquitin-like modifier (SUMO), the maltose binding protein (MBP) and the Glutathione-S-transferase (GST) that can also be used as alternative affinity tag for e.g. interaction assays. In addition, we have identified the B1 domain of the Streptococcal Protein G (GB1) as a powerful tag to increase both expression and solubility of target proteins [4], even in the case of secreted plant and fungal proteins [5-7]. Tags can be removed from the POI by a tobacco etch virus (TEV) protease cleavage site separating the POI and the respective tag(s). In addition, the backbones contain an mScarlet under control of a stationary-phase promoter allowing visual inspection of positive *E. coli* candidates in which the mScarlet expression module has been replaced by the gene of interest. This protocol further includes a section on oligonucleotide design, cloning and a detailed description of expression conditions and Ni-NTA affinity chromatography using a simple peristaltic pump. We briefly discuss a size-exclusion chromatography step to increase purity and homogeneity of the sample. Our method therefore offers an easy and rapid production strategy for both prokaryotic and eukaryotic proteins. The whole described workflow can be performed in any standard molecular biology laboratory with basic protein biochemical equipment.

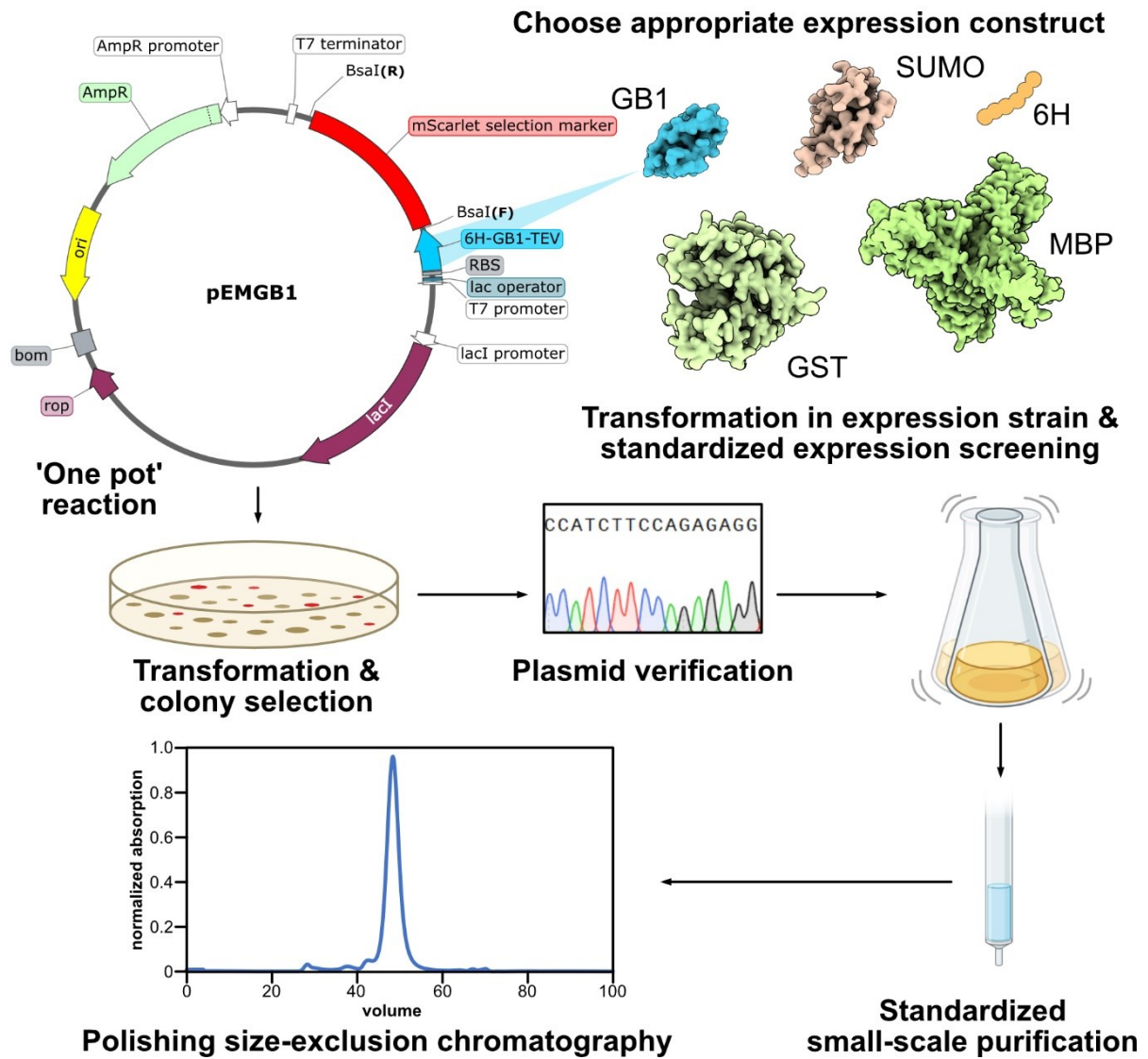

**Figure 1: Workflow for recombinant protein expression in *E. coli* using a simple Golden Gate cloning system.** Shown is the workflow from choosing the appropriate assembly system for the GOI and the one-pot-reaction into the respective expression vector. Subsequent selection and verification, followed by test expressions of the protein of interest (POI) with different methods. The protocol ends with size-exclusion step to polish the purified POI for different downstream applications. The figure was generated using Snapgene and Chimera X [8].

As an example, we use the *Thecaphora* unique effector 16 (Tue16) of the smut fungus *Thecaphora thlaspeos,* a pathogenic fungus of Brassicaceae plants.

## 2. Materials

The conduction of the described protocol requires the following standard laboratory equipment and consumables.

### 2.1. Laboratory equipment

1. Standard reagents, consumables and instrumentation for PCR reactions
2. Standard reagents for molecular biology experiments (restriction analysis, ligation)
3. Standard reagents and equipment for gel electrophoresis
4. Standard reagents, consumables and instrumentation for *E. coli* transformation.
5. Standard reagents, consumables and instrumentation for microbial cultivation
6. Standard reagents, consumables and instrumentation for plasmid extraction
7. Standard reagents, consumables and instrumentation for protein expression analysis
8. Standard reagents, consumables and instrumentation for basic protein purification
9. Peristaltic pump and FPLC system for Ni-NTA affinity chromatography and size-exclusion chromatography
10. Standard micropipettes and consumables
11. Benchtop & high-speed centrifuge
12. Ultrasonic homogenizer or cell disruption homogenizer

### 2.2. Computational equipment

A computer with standard molecular cloning software is required. In this study, the browser-based, free-to-use software Benchling® (https://www.benchling.com/) was used.

### 2.3. Plasmids

Relevant plasmids for the conduction of this protocol are listed in Table 1. Plasmids can be requested from the corresponding author.

**Table 1: Plasmids designed and used in this study.**

| Name | Relevant feature | Parental plasmid | Reference |
|---|---|---|---|
| pEMGST | Contains mScarlet to be released with BsaI; AmpR | - | This study |
| pEM6HN | Contains mScarlet to be released with BsaI; AmpR | - | This study |
| pEM6HC | Contains mScarlet to be released with BsaI; AmpR | - | This study |
| pEMGB1 | Contains mScarlet to be released with BsaI; AmpR | - | This study |
| pEMSUMO | Contains mScarlet to be released with BsaI; AmpR | - | This study |
| pEMBP | Contains mScarlet to be released with BsaI; AmpR | - | This study |
| pEMGST_Tue16 | pEMGST derivative containing Tue16 under T7 control, AmpR | pEMGST | This study |
| pEM6HN_Tue16 | pEM6HN derivative containing Tue16 under T7 control, AmpR | pEM6HN | This study |
| pEM6HC_Tue16 | pEM6HN derivative containing Tue16 under T7 control, AmpR | pEM6HC | This study |
| pEMGB1_Tue16 | pEMGB1 derivative containing Tue16 under T7 control, AmpR | pEMGB1 | This study |

### 2.4. DNA oligonucleotides

All oligonucleotides relevant for this protocol are provided in Table 2. 100 µM stocks and 10 µM working stocks are generated with $ddH_2O$ and stored at - 20 °C.

**Table 2: Listed are the oligonucleotides used in this study.**

| Name | Sequence (5´-3´) * | Information |
|---|---|---|
| Tue16_BsaI_fwd | GGTCTC**CCATGG**GCACAAACCCCCCCTCCCCTCA | Amplification of *tue16* for cloning via BsaI |

| Tue16_BsaI_rev | GGTCTC**CTCGAG**GGGGCCAGGTCCCGC | Amplification of *tue16* for cloning via BsaI |

* The color scheme is according to the explanation given in fig. 2

### 2.5. Enzymes

Other enzymes with similar properties can be applied for the conduction of this protocol. The enzymes used here were provided by New England Biolabs (NEB). (see **note 1**)
1. BsaI (10,000 units / mL)
2. Phusion DNA Polymerase (2,000 units / mL)
3. T4 DNA Ligase (400,000 units / mL)
4. TEV protease (10,000 units / mL) (if cleavage of respective tag is desired)

### 2.6. Antibiotics

The only antibiotic used in this study was ampicillin. It was stored at -20 °C in a concentration of 100 mg / mL. The final ampicillin concentration is 0.1 mg / mL.

### 2.7. Chemicals, buffers and media components

1. dYT liquid medium (1.6 % (w/v) Bacto-Tryptone, 1 % (w/v) yeast extract, 1 % (w/v) sodium chloride)
2. 10 x ligation buffer (10 mM Tris(-HCl), pH 7.5-8.0, 50 mM NaCl and 1 mM EDTA)
3. 1 x TAE buffer (1 mM EDTA · Na 2 · 2 H2O, 20 mM acetate, 40 mM Tris)
4. 6 x agarose gel loading dye
5. Agarose (standard)
6. DNA ladder (here used: ʎ/PstI)
7. YT solid medium (0.8 % (w/v) Bacto-Tryptone, 0.5 % (w/v) yeast extract, 0.5 % (w/v) sodium chloride, agar-agar 2 % (w/v))
8. 1 M isopropyl-β-D-thiogalactopyranoside (IPTG)
9. Buffers for protein purification (as stated in the following protocol)
10. Buffers for SDS-PAGE preparation (as stated in the following protocol)

### 2.8. Consumables

1. PCR reaction tubes 0.2 mL
2. Standard petri dishes
3. 1.5 mL and 2 mL reaction tubes
4. Cuvettes
5. 50 mL tube
6. 15 mL tube
7. Amicon Ultra centrifugal filters
8. 0.25 mL syringes
9. Disposable needles
10. Ni Sepharose/HisTrap FF (e.g. Cytiva)

### 2.9. Strains

The relevant laboratory *E. coli* strains for the conduction of this protocol are provided in Table 3.

**Table 3: Listed are the *E. coli* strains relevant for this study.**

| Name | Relevant features | Reference |
|---|---|---|
| K-12 TOP10 | *F- mcrA Δ(mrr-hsdRMS-mcrBC) Φ80lacZΔM15 Δ lacX74 recA1 araD139 Δ(araleu)7697 galU galK rps L (StrR) endA1 nupG* | Thermo Scientific |
| BL21 DE3 | *Δ Lon protease Δ OmpT* | [3] |

| | Δ hsdSB |  |
| | contains the T7 RNAP gene under the *lacUV5* promoter | |

## 3. Methods

### 3.1. Construction of the expression plasmid by Golden Gate cloning

#### 3.1.1. Design of oligonucleotides

This part of the method section explains the primer design on the *tue16* gene of *T. thlaspeos*. However, this approach is suitable for any gene of any target organism.

1. Use a cloning program of choice. (see **note 2**)
2. Upload the coding sequence of interest. Eukaryotic genes can contain introns. Only use the exon sequences (cDNA sequence) for oligonucleotide design. (see **note 3**)
3. Design oligonucleotides with the overhangs shown below. Do not include the codon of the starting methionine as it is already present in the 5'-overhang. (see **note 4**)

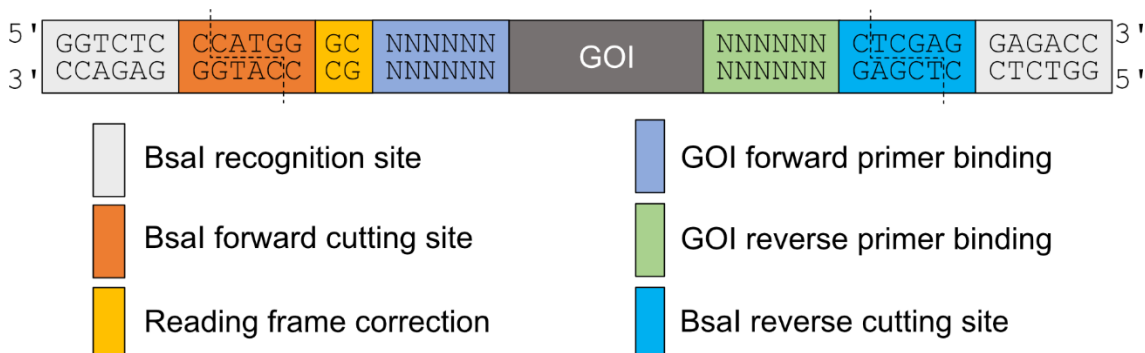

Figure 2: Schematic depiction of the Golden Gate-based cloning approach.

4. Check for similar annealing temperatures and avoid primer dimer formation using e.g. the NEB online tool (https://tmcalculator.neb.com/#!/main).

#### 3.1.2. Amplification of the GOI (see note 5)

1. If not ordered dissolved in water, apply the respective volume of ddH$_2$O to your oligonucleotides.
2. Mix the following components in a microcentrifuge tube for polymerase chain reaction (PCR):

| Constituent | Volume |
|---|---|
| ddH$_2$O | 33 µL |
| 10x Reaction buffer | 10 µL |
| gDNA/ cDNA | 1 µL |
| dNTP's | 0.5 µL |
| DNA polymerase | 0.5 µL |

3. Mix the reaction briefly by flipping the tube or pulse vortexing. Spin reactions quickly in a microcentrifuge ensuring the reaction mix is at the bottom of the tube. Place the reaction in a thermocycler running the following program:

| Temperature | Time | Cycles |
|---|---|---|
| 98 °C | 2 min | 1 |
| 95 °C | 10 - 15 s | |
| 50 – 65 °C | 10 - 30 s | 35 |
| 72 °C | 30 s for 1 kb | |
| 72 °C | 5 min | 1 |
| 4 °C | hold | 1 |

4. Prepare an agarose gel for subsequent gel electrophoresis. Weigh in the respective amount of agarose, mix with the respective volume of 1x TAE buffer and heat it in a microwave oven for 2-5 minutes until the solution appears translucent. Let it cool to approximately 60 °C and apply your DNA stain in the appropriate concentration. Pour the liquid in a leakproof gel tray and let it solidify for about 10-15 minutes. Place gel tray in a chamber filled with 1x TAE buffer so that it is covering the gel completely (see **note 6**).
5. Apply PCR samples onto the wells of the gel previously mixed with 6x loading dye buffer and apply a DNA size standard.
6. Apply 120 V for 45 minutes and take a photo of stained DNA with an appropriate imager system.
7. Cut out a gel slice at the respective size with a scalpel and transfer it to a 2 mL reaction tube. Wear protective equipment when working under UV light.
8. Use an appropriate gel extraction kit and subsequently measure the DNA concentration using a UV-Vis spectrophotometer (e.g. NanoDrop, ThermoScientific).

### 3.1.3. Golden Gate Cloning

1. Determine concentrations and prepare all DNA fragments relevant for the Golden Gate reaction (see **note 7**)
2. Mix the following reagents in a PCR reaction tube (see **note 8**):

| Constitutent | Volume |
|---|---|
| T4 DNA Ligase buffer | 1 µL |
| T4 DNA Ligase (400,000 units / mL) | 1 µL |
| Type IIs restriction enzyme (here: BsaI) | 1 µL |
| Destination plasmid | ~20 fmol |
| GOI fragment | ~20 fmol |
| In case of annealed oligonucleotides | 20 fmol up to 2 pmol |
| ddH$_2$O | Ad 10 µL |

3. Mix the reaction briefly by flipping the tube or pulse vortexing. Spin reactions quickly in a microcentrifuge ensuring the reaction mix is at the bottom of the tube. Place the reaction in a thermocycler running the following program:

| Temperature | Time | Cycles |
|---|---|---|
| 37 °C | 2 min | 10 |
| 16 °C | 5 min | |
| 37 °C (see **note 9**) | 5 min | 1 |
| 16 °C | 5 min | 1 |
| 80 °C | 5 min | 1 |
| 16 °C | Hold | 1 |

4. Use the Golden Gate reaction immediately for *E. coli* transformation or store at -20 °C for later transformation (see **note 10**).

### 3.1.4. Transformation of competent *E. coli* cells

For this protocol, in-house generated competent *E. coli* TOP10 cells were used. Chemically competent cells were prepared according to a RbCl-based method [9]. Nonetheless, other suitable methods for the preparation of competent *E. coli* cells can be used as well.
1. Thaw 20 µL of competent cells for 5 minutes on ice.
2. Add 5 µL of the Golden Gate reaction mixture and mix by gently snipping a finger against the microcentrifuge tube (see **note 10**).
3. Incubate on ice for 5 minutes, while heating a water bath or heat block to 42 °C.
4. Perform a heat shock at 42 °C for 45 seconds and immediately put the microcentrifuge tube back on ice afterwards.

5. After 5 minutes, add 300 µL of your preferred medium. (Ideally preheated to 37 °C.)
6. Incubate at 37 °C at a shaking frequency of 200 rpm for 30 minutes
7. Plate 150 – 300 µL of the cell suspension on agar plates supplemented with the medium of preference and ampicillin using sterile glass beads/ pipettes or a cell spreader.
8. Incubate plates over night at 37 °C.

### 3.1.5. Screening and plasmid verification

An mScarlet expression module inserted between the BsaI sites in the used vector set allows for visual inspection of transformants and identification of potentially positive candidates. A *glpT* promoter that is induced by low glucose concentrations (i.e., in stationary phase) is used for *mScarlet* gene expression based on the iGEM part BBa_J72163. Red colonies producing the mScarlet can be clearly differentiated from the potential positive 'white' ones. In this protocol, we extract plasmid DNA and use a restriction digest followed by external sanger sequencing to confirm the correct DNA sequence.

1. Inoculate potentially correct 'white' colonies (2 – 5 colonies; see **note 11**) in 5 mL dYT media with ampicillin and incubate over night at 37 °C under continuous shaking at 200 rpm.
2. Purify plasmid DNA with a method of choice.
3. Verify the correct insert by choosing appropriate restriction enzymes for a restriction digest. (see **note 12**)
4. Send the approved candidates for Sanger sequencing using an appropriate sequencing primer.

## 3.2. Protein expression of the target protein

In this protocol we only describe protein production in *E. coli* BL21 DE3. Depending on the produced protein, other expression strains should be chosen. (see **note 13**)

### 3.2.1. Transformation of *E. coli* expression strain

1. To transform the respective *E. coli* strain with the verified plasmid, thaw 20 µL of the respective competent *E. coli* cells on ice for 5 minutes for each protein of interest.
2. Add 1 µL of the final plasmid (concentration 50 - 150 ng/µL) onto the thawed cells. Mix carefully by flipping your finger against the reaction tube. Proceed as described in 3.1.4.

### 3.2.2. Test expression of the target protein

1. For test expression, three different conditions are tested. Of each culture, cell material of different colonies is gathered by wiping over the agar plate containing the freshly transformed expression strain with a sterile glass pipette. (see **note 14**)
2. For the autoinduction, inoculate your culture(s) of dTY-medium with appropriate antibiotic concentration and 1 % (w/v) lactose (**Lac**) and incubate the culture at 28 °C for 20 hours at 180 – 220 rpm.
3. For IPTG induction, inoculate pre-cultures (1/10 of your main culture volume) and after overnight incubation (12 to 16 h) upon shaking at 180 – 220 rpm, dilute to an optical density (OD600) of 0.1 – 0.2. Incubate the cultures until they reach an optical density of 0.5 – 0.7. One of the cultures is induced with 0.5 mM IPTG and incubated at 37 °C for 3 h at 180 – 220 rpm (**IPTG37**). The other culture needs to be cooled until it reaches 20 °C and is then induced with 0.5 mM IPTG to be incubated at 20 °C for 20 h, shaking at 180 – 220 rpm (**IPTG20**).

4. After the respective incubation time, harvest cells at 4.000 x *g* for 15 minutes. (see **note 15**) Either freeze cells with liquid nitrogen and store at – 20 °C or continue with the subsequent small-scale protein purification protocol.

### 3.3. Small-scale protein purification

For a small-scale protein production (culture volumes < 0.05 L), the following buffers should be prepared a priori. Each buffer should be adjusted to pH 8.0 and filtered through a 0.45 μm pore size filter. (see **note 16**)

| Name | Components |
|---|---|
| Buffer A | 20 mM HEPES, pH 8.0<br>20 mM KCl<br>40 mM Imidazole<br>250 mM NaCl |
| Buffer B | 20 mM HEPES, pH 8.0<br>20 mM KCl<br>500 mM Imidazole<br>250 mM NaCl |
| 3 x Laemmli buffer | 150 mM Tris-HCl pH 6.8<br>6 % (v/v) SDS<br>30 % (v/v) Glycerol<br>15 % (w/v) ß-Mercaptoethanol<br>0.003 % (w/v) Bromphenol blue |

All following steps are conducted on ice and the respective buffers have to be cooled to 4 °C prior to use.

1. Resuspend harvested cells in 10 mL ice cold buffer A to continue with the subsequent protein purification.
2. For cell lysis, prepare a 10 mg / mL lysozyme solution (dissolve in water or buffer A) and add 100 μl per 10 mL of culture (100 μg / mL final concentration). Sonicate each cell suspension for 5 x 30 s using 70 % power (see **note 17**) and take a sample of 20 μL into a new reaction tube (**C** - cell sample). Add 10 μL 3 x Laemmli buffer to the sample.
3. Centrifuge the cells for 20 min at 16.000 x *g* at 4 °C to separate cell debris from the lysate, transfer the supernatant to a fresh 15 mL tube. Take 20 μL as a sample from the supernatant (**L** - lysate sample) and mix it with 10 μL 3 x Laemmli buffer.
4. Add 50 μl of Ni-NTA agarose (shake gently to resuspend the agarose resin before use) with a cut pipette tip and incubate at 4°C for 15 - 20 min under continuous rotation. (see **note 18**).
5. Centrifuge the samples at 1.500 x *g* and 4 °C for 5 min (see **note 19**) and carefully discard the supernatant (do not touch the sedimented resin).
6. Add 1 mL of buffer A and transfer the resin to a fresh 2 mL reaction tube and wash the pellet thrice with 1 mL Buffer A (centrifuge and remove the supernatant after each wash). After the last wash, remove the supernatant. For elution, add 50-100 μL buffer B and gently invert the tube to mix. After incubation for 5 min at 4 °C, centrifuge at 1.500 x *g* for 5 minutes.
7. Take a 20 μL sample from the supernatant (eluate sample) and mix with 10 μL 3 x Laemmli buffer (**E** – elution sample).
8. For analysis, prepare an SDS gel. First, prepare the resolving gel, pour the mixture into an appropriate gel chamber and add 2-propanole until the surface of the gel is even. After polymerization, remove the leftover 2-propanole and add the freshly prepared stacking gel mixture. Immediately append the appropriate comb into the stacking gel. (see **note 20**)

| SDS-PAGE gel | Components |
|---|---|
| Resolving gel | 0.375 M Tris-HCl, pH 8.8<br>10 % (v/v) Acrylamide |

| | 0.1 % (v/v) SDS<br>1.25 % (v/v) Glycerol<br>0.05 % (v/v) Ammonium persulfate (APS)<br>0.1 % (v/v) Tetramethylethylenediamine (TEMED)<br><br>In $H_2O$ (bidest.) |
|---|---|
| Stacking gel | 0.125 M Tris-HCl, pH 6.8<br>5 % (v/v) Acrylamide<br>0.1 % (v/v) SDS<br>0.05 % (v/v) Ammonium persulfate (APS)<br>0.1 % (v/v) Tetramethylethylenediamine (TEMED)<br><br>In $H_2O$ (bidest.) |

9. For analysis, load the three samples of each protein expression (**C**ell, **L**ysate, **E**luate) onto an SDSgel with a protein ladder appropriate to the size of your POI. Let it run at a constant voltage of 250 V for 30-40 minutes and stain with a Coomassie Blue staining solution for 30 minutes. (see **notes 21 and 22**) Destain the gel by adding destaining solution multiple times until the background dye is removed.

| Name | Components |
|---|---|
| Coomassie brilliant blue-R250 staining solution | 0.05 % (w/v) Coomassie Brilliant Blue R250 (CBB)<br>9 % (v/v) Glacial acetic acid<br>45 % (v/v) Methanol<br><br>In $H_2O$ (bidest.) |
| Destaining solution | 10 % (v/v) Glacial acetic acid<br>10 % (v/v) Methanol<br><br>In $H_2O$ (bidest.) |

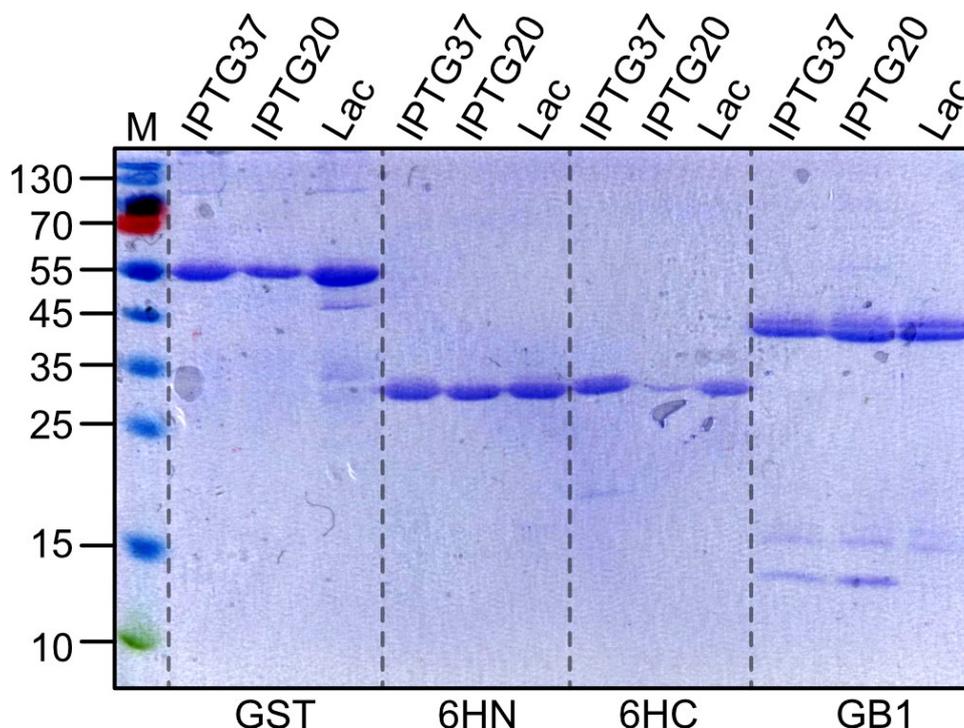

**Figure 3. Expression and purification of tagged Tue16 from different backbones using a predefined set of expression conditions.** The polyacrylamide gel shows the elution fractions of Tue16 with four different tags expressed under the described three different standard conditions (1=IPTG37;2=IPTG20;3=Lac). In case of Tue16, only slight differences in purified protein amounts are observed using the different expression conditions.

### 3.4. Large-scale protein purification

After evaluating optimal expression conditions providing the highest yields of your POI in a small-scale expression (**see chapter 3.3**), a large-scale protein purification can be performed to obtain large yields of highly pure protein for biophysical, biochemical or structural analysis. Follow the instructions of the respective expression method in chapter 3.3 but increase the culture volume up to 6 L and elute the harvested cells in 30 - 35 mL Buffer A per liter of expression culture. For the large-scale protein purification, you will typically need 1 L of Buffer A, at least 50 mL of Buffer B and 1 L of SEC Buffer. Each Buffer should be previously filtered through a 0.45 µm filter. All steps are conducted on ice and all buffers are cooled prior to use.

| Name | Components |
|---|---|
| SEC buffer | 20 mM HEPES, pH 7.5<br>20 mM KCl<br>200 mM NaCl |

#### 3.4.1. Cell lysis

1. After cultivation, cells have to be harvested in a centrifuge with the respective capacity for 15 min at 4.000 x *g*. Eventually freeze the pellet at -20 °C until further use.
2. Take the harvested or thawed cells and resuspend them in 30-35 mL Buffer A per liter of culture.
3. Meanwhile, cool the centrifuge to 4 °C with the respective rotor (here: JA25.50) inside.
4. For cell lysis of larger culture volumes, a microfluidizer is used (see **note 23**). Flush the microfluidizer at least twice with cooled Buffer A and pour in the resuspended sample through a sieve to avoid plugging the machine. Start disrupting the cells but discard the first bit of flow-through of buffer A as it does not contain your sample yet. Collect the disrupted cells and repeat once more but collect everything. Wash the system with up to 10 – 15 mL Buffer A and collect the flow-through to avoid discarding some of the disrupted cells. (see **note 24**)
1. Transfer the lysed cells into 50 mL centrifugation tubes, take a sample (**C** – cell sample) and balance them out on a scale up to two decimal places by adding cooled Buffer A to the lighter sample. Transfer them into the JA25.50 rotor of the centrifuge.
2. Centrifuge for 20 minutes at 4 °C at 40,000 x *g*.
3. In the meantime, calibrate your protein purification system for the later Size-Exclusion Chromatography (SEC) by connecting the respective column and starting the system equilibration. (see **note 25**)
4. Also check your peristaltic pump and start equilibration of your Ni-NTA columns that you want to use. For that, connect the Ni-NTA columns to the peristaltic system and equilibrate with 10 column volumes (CV) pre-cooled Buffer A. Depending on the flow rate, the equilibration approximately takes 10-20 minutes. (see **notes 26 and 27**)

#### 3.4.2. Ni-NTA purification

1. After the centrifugation, transfer the supernatant of each protein sample into a new 50 mL tube and take a sample (**L** – Lysate sample). (see **note 28**)
2. Stop the peristaltic pump, transfer the tubing into the tube containing the cleared lysate and start the pump. Put a new beaker underneath the opening of the columns to collect the flow-through. Collect a sample of the Flow-through (FT– flow-through) by transferring 20 µL into a new reaction tube and adding 10 µL of 3 x Laemmli buffer.
3. Wash the columns for 5-10 CV with buffer A and collect the wash in a new beaker. Take a sample of 20 µL of the wash and add 10 µL of 3 x Laemmli buffer (**W** – wash sample).

4. Disconnect the column(s) from the tubing and flush the tubing system with buffer B. (see **note 29**)
5. Reconnect the column(s) to the tubing, place a fresh 50 mL tube under the columns elute your protein of interest in 5 CV.
6. Take an elution sample (**E** – elution sample) by transferring 20 µL of the eluate into a reaction tube and adding 10 µL of 3 x Laemmli buffer.

### 3.4.3. Reverse-Ni-NTA purification *(optional)*

For the cleavage of the respective tag from your protein of interest, His-tagged TEV protease is used. If no cleavage is planned, continue with section 3.4.4.

1. Transfer the eluate into two appropriate Amicon concentrators. This depends on the size of your fusion protein. Balance out the Amicon concentrators using a precision scale. (see **note 30**)
2. Centrifuge at 5.000 – 7.500 x *g* with a fixed angle rotor or at 4,000 x *g* with a swing-out rotor at 4 – 8 °C until the remaining volume reaches the 1-0.5 mL mark. This can take 30 minutes to several hours.
3. Fill the concentrator with precooled SEC buffer that does not contain imidazole and concentrate again to 1-0.5 mL.
4. Fill the concentrator up again to the 15 mL mark with SEC Buffer and add an appropriate amount of TEV (use 1 OD of TEV for 100 OD of your protein; efficiency might be different for different proteins). Incubate for 2 – 3 h at room temperature or overnight at 4 °C while shaking on a rotary wheel or rocker.
5. Load your TEV-cleaved sample on a regenerated Ni-NTA column and take a sample by transferring 20 µL of the probe in a reaction tube and adding 10 µL of the 3 x Laemmli buffer (**TE** – TEV-cleaved sample).
6. Start the peristaltic pump to load your protein onto the column. Collect the flow-through in a fresh 50 mL tube (see **note 31**). Take a 20 µL sample and mix it with 10 µL 3 x Laemmli buffer (**TF** – TEV-cleaved flow).
7. Wash the column with 5 – 10 CV of Buffer A and collect it in a new 50 mL tube (see **note 31**). Take a 20 µL sample and mix it with 10 µL 3 x Laemmli buffer (**TW** – TEV-cleaved wash).
8. Elute the cleaved tag and TEV protease with 5 CVs and collect it as well in a fresh 50 mL tube. Take a 20 µL sample and mix it with 10 µL 3 x Laemmli buffer (**TE** – TEV-cleaved elution).

### 3.4.4. Sample analysis & preparation for size-exclusion chromatography

1. Transfer the eluate into an appropriate Amicon concentrator. This depends on the size of your protein construct. Balance out the Amicon concentrators using a fine scale. (see **note 30**)
2. Centrifuge at 5,000 – 7,500 x *g* with a fixed angle rotor or up to 4,000 x *g* with a swing-out rotor at 4 – 8 °C until the remaining volume reaches approx. 1-0.5 mL. This can take 30 minutes to several hours depending on protein concentration. (see **note 32**)
3. Meanwhile, load 5 µL of the collected lysate, flow-through, wash, elution sample and potentially samples from the reverse-Ni-NTA onto an SDS-PAGE, adding 5 µL of an appropriate protein ladder into an additional lane and let it run at a constant voltage of 250 V for 30-40 minutes. Image your SDS-gel using a UV-imaging system and take a picture to check if your elution fraction contains your desired fusion protein. (see **note 33**)
4. Transfer the separating gel into an air-sealed box, cover the gel with Coomassie-Blue solution and incubate it on a rocker for at least 30 minutes. After incubation, remove the Coomassie-Blue solution and add destaining solution to the box, covering the gel completely. Change the destaining solution 3 – 5 times, each after at least 30 minutes of incubation. Once the gel is fully destained, use an imaging system to take a picture for documentation. (see **note 34**)
5. For regeneration of the Ni-NTA columns, wash them with ddH$_2$O, followed by 0.5 M EDTA solution, 0.5 M NaOH and 0.2 M NiSO$_4$, with ddH$_2$O washing steps in between and at the end for 5 CVs each. (see **note 35**)
6. Once the elution fraction has been concentrated to 1 - 0.5 mL, resuspend the concentrated protein solution and transfer into a cooled 1.5 mL Reaction tube.
7. Clean the concentrator by filling it with SEC Buffer and subsequent centrifugation until about 1 mL of the Buffer is left on the membrane. Used, clean concentrators can be stored in H$_2$0. (see **note 36**)

### 3.4.5. Size-exclusion chromatography

This protocol aims at providing a simple and convenient methodology for protein purification. Size-exclusion chromatography is the method of choice for polishing purified proteins and retrieving homogenous samples. However, for the sake of simplicity, we will not introduce pros and cons of choosing appropriate columns and buffer conditions here but rather give general advice on protein handling.

1. Spin down your concentrated protein at 8.000 x *g* for 1 minute to remove precipitants and measure the absorption at 280 nm to determine the protein concentration. (see **note 37**)
2. Wash the injection loop of the FPLC-system with 10 loop volumes of SEC buffer using a syringe. (see **note 38**)
3. Take up your concentrated protein solution with a syringe and inject into the injection loop.
4. Run an appropriate program depending on the choice of columns and the used buffer conditions. Collect the elution fractions using a fraction collector.

### 3.4.6. Sample analysis

1. After the run has finished, open the respective evaluation program and choose up to 14 peak fractions for analysis (depending on the available SDS-PAGE setup), Of each fraction, take a 20 µL sample and mix it with 10 µL of 3 x Laemmli-Buffer. Run an SDS-PAGE at 250 V for 40 minutes with an appropriate protein size standard. Proceed with the SDS-PAGE as stated in section 3.3. Once the SDS-PAGE is documented, protein purity can be evaluated, and appropriate fractions are chosen for concentration.
2. Pool the fractions containing your pure protein into one or several Amicon-concentrators and centrifuge at 5,000 – 7,500 x *g* with the fixed angle rotor or at 4.000 x *g* using a swing-out rotor at 4 – 8 °C until the volume reaches 0.5 - 1 mL. Measure the absorption at 280 nm to determine the protein concentration.
3. Prepare aliquots of your POI according to downstream analysis purposes and snap freeze in liquid nitrogen for long-term storage at -80 °C. (see **note 39**)

### 3.5. Summary & Perspectives

In conclusion, we here present an efficient and easily accessible approach to recombinant protein production that combines Golden Gate cloning with streamlined protein purification. It therefore might enable especially researchers from laboratories not specialized in biochemistry to explore protein functions and interactions. Researchers may also explore additional solubility tags and affinity purification strategies further than the ones presented here. In the future, we anticipate incorporating an additional set of plasmid backbones with different resistance markers and origins of replications to allow co-expression of two or more proteins to enhance the versatility of the system with respect to hetero-oligomeric complexes.

## 4. Notes

1. Enzymes should be stored at -20 °C and rigorously kept on ice during use.
2. In this study, the browser-based, free-to-use laboratory notebook Benchling® (https://www.benchling.com/) was used that also features a comprehensive molecular cloning tool. However, any cloning program of choice can be used.
3. It is advisable to retrieve sequences from the respective genomic database of your organism of choice, if available. These databases can often be accessed via the UniProt database (https://www.uniprot.org).
4. Make sure to correctly design the overhangs to avoid a frame shift in the open reading frame. The overhangs include the codon ATG for starting methionine and a GC has to be added in the oligonucleotide sequence to correct the reading frame and bring the GOI *in frame* with the respective fusion tag.
5. In this protocol, the gene of interest was amplified from genomic DNA. However, it is possible to use other scaffolds for Golden Gate cloning such as (i) annealed oligonucleotides or (ii) synthesized DNA fragments with the respective overhangs.
6. We use a standard agarose concentration of 1% (w/v).
7. Using e.g. a nanodrop to determine the absorption at 260 nm is usually sufficient.
8. Enzymes should be kept at -20 °C and added last.
9. If the gene of interest has internal BsaI recognition sites, this step has to be omitted. The overall efficiency of the Golden Gate reaction will decrease but chances are your correctly assembled plasmid can be obtained.
10. The remaining reaction can be stored at -20 °C and used for another transformation with a larger volume in case of low transformation efficiency.
11. The maturation of the mScarlet can sometimes take up to 24 h. If distinguishing between white and red colonies is difficult due to pale red color, storage at 4 °C up to 24 h is recommended. The positive rate is usually high (< 10% red colonies) but can vary in case of long fragments or the assembly of an open reading frame from several fragments.
12. Ideally, choose a single enzyme that digests the backbone in 2-4 fragments that can be clearly distinguished by size and are not smaller than 500 bp for optimal resolution. The restriction pattern should be different from the hydrolysis of the ancestor plasmid. Often, molecular cloning programs offer options for virtual digestion.
13. In our hands, BL21 DE3 retrieves/provides good results for the majority of proteins produced. Especially in the case of proteins with disulfide bonds, we made good experience with using SHuffle T7 (NEB).
14. Always transform the expression strains freshly prior to performing expression tests. Transformation plates should not be older than one week.
15. Centrifugation time depends on the culture volume. A time of 15 minutes is sufficient to harvest culture volumes up to 1 L.
16. Buffers can be stored for several days to weeks at 4 °C. Filtration extends the shelf life of buffers and removes contaminations and non-dissolved precipitants. Carefully inspect your buffers prior to use.
17. The given values have to be adjusted to different ultrasonic homogenizer models. It is important to wear noise protective equipment in order to avoid long-term hearing damage.
18. An incubation time of 15-20 minutes is sufficient to ensure optimal protein binding to the Sepharose resin in our hands. However, the time can be extended or even shortened to optimize the results.
19. According to manufacturer's instructions, the sedimentation of Ni Sepharose is usually performed at 500 x *g* for 5 min. However, we did not observe any issues in bead stability or protein binding by using higher *g* forces.
20. SDS gels can be stored at 4 °C without the combs and wrapped in wetted paper towels for several weeks. In case of regular use, we recommend batch preparation of up to 12 gels to ensure homogenous separation properties and quality.

21. Running the polyacrylamide gels at 250 V is only possible when using a maximum of two gels as the increased current might lead to overheating. Always use fresh buffer to allow an equal electrolyte separation.
22. A stain-free approach visualizing separated proteins using UV light is possible when adding 2,2,2-trichloroethanol (0.5 % final concentration) during polyacrylamide gel preparation [10].
23. An ultrasonic homogenizer can also be used for larger culture volumes by increasing homogenization cycles and time. However, if access to a cell disruptor is available, we always recommend this cell lysis method due to lower temperatures and a faster cell lysis preventing protein denaturation or aggregation.
24. Immediately continue with the next step to avoid degradation of your POI.
25. The time for column equilibration varies depending on the used column and the flow rates. We typically store our size-exclusion columns in filtered $H_2O$ due to frequent (daily) use. Equilibration with 1 ½ CV of SEC Buffer therefore only takes approx. 2 hours. In case of a less frequent use, long-time storage in 20 % (v/v) ethanol increases the column lifetime.
26. We recommend using a 1 ml Ni-NTA column per liter of *E. coli* culture. However, depending on protein expression, usage of a 5 ml column might also be appropriate. Of note, Ni-NTA FF columns can also be stacked (e.g. 1 ml + 1 ml columns).
27. Ideally, the peristaltic pump should be located in a cooling cabinet or cold room (adjusted to 4 °C). Note the abbreviation on the columns: HP – HighPerformance (should not be stacked), FF – FastFlow (stackable). Avoid pumping air on the columns.
28. Only transfer cleared lysates without disturbing the cell debris pellet. In case of remaining debris, repeat the centrifugation step. Any remaining particles can potentially clog the Ni-NTA columns.
29. Put the tubing right to the bottom of your tube and stop the pump when only about 1 mL of lysate is left in the tube to avoid filling the tubing with air. We recommend measuring the exact flow rate and setting a timer for all upcoming steps as well on the peristaltic pump to minimize the risk of letting the columns run dry.
30. To calculate the molecular weight of your fusion protein we recommend using the ProtParam tool (https://web.expasy.org/protparam/). Do not use a 50 mL tube tube to balance out the Amicon concentrator in the centrifuge since it has a different center of gravity. Use multiple concentrators for one protein construct to reduce centrifugation time. Whilst centrifugation, check the concentrators for leakage and balance them out once more to avoid imbalance of the centrifuge.
31. Your protein will most likely not bind to the column anymore and should be present in the flow-through. However, depending on the remaining concentration of imidazole in your buffer, it might also elute upon washing the column in the next step.
32. The appropriate end-volume of the concentration has also to be adjusted depending on protein stability and on the used injection loop (0.5 – 5 mL). The solubility limit of a given protein might be reached prior to this point, resulting in precipitation of the sample. In any case, we recommend using the smallest injection loop possible and stick to the manufacturer's instructions for the respective SEC column. Injection of large volumes on small columns will influence the separation on the SEC column.
33. Wear gloves while the handling of the SDS-PAGE, as well as the staining and destaining of the gel due to unpolymerized acryl-amid. Use an UV-detectable protein ladder for quick analysis under UV light. (e. g. not a prestained size standard). Note that the UV detection relies on a photoreaction of tryptophan with trichloroethanol and proteins that do not contain tryptophan residues will not be detectable.
34. Depending on the purpose, UV documentation might be sufficient. For quicker Coomassie-Blue staining and destaining, carefully heat the gel with respective solution for 30 s in a microwave and afterwards, incubate it for 20 more minutes on a shaker. You can collect the used Coomassie-Blue solution for re-use.

35. Collect the waste fraction containing NiSO$_4$ separately and dispose appropriately.
36. Amicon spin concentrators can be re-used several times. Cleaning with 0.2 M NaOH allows to remove precipitated protein and cleans the membrane. For storage, we suggest to fill the concentrators completely with sterile H$_2$0.
37. To calculate the protein concentration on the basis of the measured absorption at 280 nm, use the following calculation: A = ε x c (A: measured absorption; ε: extinction coefficient of the protein; c: concentration in mmol/ml)
38. To prevent column clogging, we recommend to test protein stability in the used SEC buffer prior to application. Therefore, before loading a newly produced protein onto the SEC column, mix 50 µL of concentrated protein with 50 µL SEC buffer and check for precipitation. In case of visible precipitation, optimize the SEC Buffer accordingly (e.g. by adding more salt or glycerol). As buffer optimization can be time consuming and tedious, we will not cover this topic here.
39. If working with enzymes, up to 50 % glycerol might be added prior to snap freezing. For most other proteins, snap freezing and storage at -80 °C does not require the addition of glycerol.

**Acknowledgements & Statement**

We thank Alexander Lepak for his efforts on introducing an earlier version of this Golden Gate approach and critical discussions. We also acknowledge Paul Weiland for providing the pEMSUMO and pEMBP backbones. FA and SZ acknowledge funding through the German Research Foundation (Deutsche Forschungsgemeinschaft, DFG) – (Project ID 458090666 / CRC1535/B02). The research was supported by the German Research Foundation (Deutsche Forschungsgemeinschaft, DFG) *via* the Collaborative Research Centre 1208 "Identity and Dynamics of Membrane Systems—from Molecules to Cellular Functions" (to FA). All materials are available from the corresponding author upon request.

**Author contributions**

FA designed the study. FA and SZ wrote the paper with input from all authors. SZ and GM performed experiments and analyzed data. FA contributed funding and resources.